\documentclass[10pt,draft]{dis03}
\usepackage{epsf,amsmath}

\begin{document}

\title{Quark distributions in polarized and nonpolarized hadrons \\
PREPARED BY LATEX
 \thanks{}}

\author{DIS~0.~3.~ A.G. Oganesian\\
Institute of Theoretical and Experimental Physics,\\
Moscow,117218,Russia\\
E-mail: armen@heron.itep.ru}

\maketitle

\newcommand{\be}{\begin{equation}}
\newcommand{\ee}{\end{equation}}

\def\la{\mathrel{\mathpalette\fun <}}
\def\ga{\mathrel{\mathpalette\fun >}}
\def\fun#1#2{\lower3.6pt\vbox{\baselineskip0pt\lineskip.9pt
\ialign{$\mathsurround=0pt#1\hfil##\hfil$\crcr#2\crcr\sim\crcr}}}

\begin{abstract}
\noindent Valence quark distributions in pion,  polarizied $\rho$
mesons and nucleon in the region of intermediate $x$ are obtained
by generalizied QCD sum rules. It is shown, that polarization
effects are very significant. The strong suppression of quark and
gluon sea distributions in longitudinally polarizied $\rho$ mesons
is found.
\end{abstract}

Determination of quark  distribution functions in a
model-independent way in QCD sum rules based only on QCD and the
operator product expansion (OPE) seems to be very important task
especially for polarized hadron. A method to determinate valence
quark distribution in QCD sum rules at intermediate $x$ was
suggested in \cite{1} and generalized in \cite{2}-\cite{3}. Let me
briefly present the method. We start from consideration of the
4-point correlator with 2 hadron and 2 electromagnetic currents.

%3
$ \Pi = -i \int~ d^4x d^4y d^4z e^{ip_1x + iqy - ip_2z} \langle 0
\vert T \left \{j^h(x),~ j^{el}(y),~ j^{el}(0),~ j^h(z) \right \}
\vert 0 \rangle $

Here $p_1$ and $p_2$ are the initial and final momenta carried by
hadronic the current $j^h$, $q$ and $q^{\prime} = q + p_1 - p_2$
are the initial and final momenta carried by virtual photons
(Lorentz indices are omitted) and $t = (p_1 - p_2)^2 = 0$. To find
structure function one should compare dispersion representation of
the forward scattering amplitude in terms of physical states with
those in OPE and use Borel transformation. Though we should
consider the case of forward scattering, i.e. $p_1=p_2$, but, and
it is significant to method, we should keep $p_1^2$ not equal to
$p_2^2$ in all intermediate stages and only in final result take
the forward scattering limit. Only in such way, and it was the
main idea of generalization of sum rules (see \cite{2}) it was
found to effectively (exponentially) suppress all terms in sum
rules expect those which are proportional to structure function so
we can found structure functions of hadron with good accuracy.
Finally equating the physical and QCD representations (for detail
see \cite{2}) after double borelization (on $p_1^2$ and $p_2^2$)
we found

%10
$
Im~ \Pi^0_{QCD} + \mbox{Power~ correction} = 2 \pi F_2 (x, Q^2) g^2_h e
^{-m^2_h(\frac{1}{M^2_1} + \frac{1}{M^2_2})}
$

Here $Im \Pi^0_{QCD}$ is the bare loop contribution, (where
continuum contribution is eliminated)and $g_h$ is defined as
$\langle 0 \vert j_h \vert h \rangle = g_h $. In what follows, we
put $M^2_1 = M^2_2 \equiv 2 M^2$.

Let me briefly illustrate the main points of the calculation for
the case of pion. It can be treated as a check of the accuracy of
the method due to fact that for pion the experimental results are
available. To find the pion structure function one should choose
the imaginary part of 4-point correlator  with two axial (hadron)
and two electromagnetic currents and consider the invariant
amplitude at tensor structure, $P_{\mu} P_{\nu} P_{\lambda}
P_{\sigma}/\nu$, where $P=(p_1+p_2)/2$, ${\mu},{\nu}$ are vector
current indexes and ${\lambda}, {\sigma}$ -hadron current indexes.
I shall briefly note the main points of calculations, for detail
see \cite{2}. In QCD part of sum rules we take into account the
following terms:

1. Unit operator contribution (bare loop) and leading order (LO)
perturbative corrections proportional to $ln(Q^2/\mu^2)$ , where
$\mu^2$ is the normalization point. In what follows, the
normalization point will be chosen to be equal to the Borel
parameter $\mu^2 = M^2$.

2.  Power corrections -- higher order terms of OPE. Among them,
the dimension-4 correction, proportional to the gluon condensate
$\langle 0 \vert \frac{\alpha_s}{\pi} G^n_{\mu \nu}~ G^n_{\mu\nu}
\vert 0 \rangle$ was first taken into account, but it was found
that the gluon condensate contribution to the sum rule vanishes
after double borelization. There are two types of vacuum
expectation values of dimension 6. One, involving only gluonic
fields: $\frac{g_s}{\pi} \alpha_s f^{abc} \langle 0 \vert
G^{a}_{\mu \nu}~ G^b_{\nu \lambda}~ G^c_{\lambda \mu} \vert 0
\rangle$  and the other, proportional to the four-quark operators
(after factorization) $a^2=\alpha_s (2\pi)^4 (\langle 0 \vert
\bar{\psi} \psi \vert 0 \rangle)^2 $. Terms of the first type
cancel in the sum rule for pion and only terms of the second type
survive. Finally quark distribution function has the following
form:

%19
$$
xu_{\pi}(x) = \frac{3}{2\pi^2}\frac{M^2}{f^2_{\pi}}x^2(1-x)
\Biggl [ \Biggl ( 1+ \Biggl (\frac{a_s(M^2)\cdot ln(Q^2_0/M^2)}{3\pi}\Biggr
)$$

%$$\times \Biggl ( \frac{1+4x ln(1-x)}{x}- \frac{2(1-2x)ln
%x}{1-x}\Biggl ) \Biggl )\cdot N(x)(1-e^{-s_0/M^2}) $$

\be
\times N(x)(1-e^{-s_0/M^2})-\frac{4\pi \alpha_s(M^2)\cdot 4\pi
a^2}{(2\pi)^4 \cdot 3^7\cdot 2^6\cdot M^6} \cdot
\frac{\omega(x)}{x^3(1-x)^3}\Biggr ], \label{19} \ee where
$N(x)=(1-x)(1+4xln(1-x))-2x(1-2x)lnx~$ and $\omega(x)$ is the
fourth degree polynomial in $x$,
 This function
$u_{\pi}(x)$ may be used as an initial condition at $Q^2 = Q^2_0$
for solution of QCD evolution equations.

\begin {figure}[!thb]
\vspace*{1.0cm}
\begin{center}
%\special{psfile=dis03_template_Fgs3.eps voffset=-60 vscale=40
%hscale= 40 hoffset 10 angle=0}
\centerline{\epsfxsize=2.9in\epsfbox{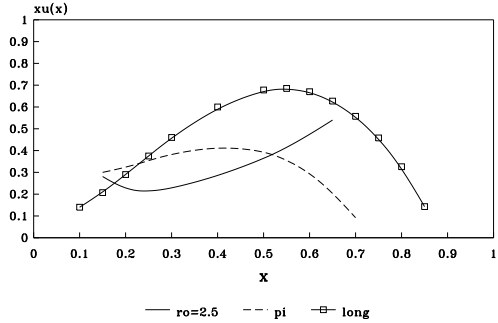}} \caption[*]{Quark
distribution function for pion, transversally and longutuninally
polarized $\rho$ - dash, solid and line with squares
correspondingly}
\end {center}
\end {figure}

In the numerical calculations we choose: the effective
$\Lambda^{LO}_{QCD} = 200 ~MeV$, $Q^2_0 = 2 ~GeV^2$. $a^2$ is
varied $a^2=0.23\pm 0.1 GeV^6$, the upper limit is close to those
found from recently from $\tau$ decays analysis \cite{4}, the
middle is close to standard choose. The continuum threshold was
varied in the interval $0.8 < s_0 < 1.2GeV^2$ and it was found,
that the results depend only slightly on it's variation. The
analysis of the sum rule (\ref{19}) shows, that it is fulfilled in
the region $0.2 < x < 0.7$, where both power correction and
continuum contribution are less than $30\%$ and the stability in
the Borel mass parameter $M^2$ dependence in the region $0.4 GeV^2
< M^2 < 0.7 GeV^2$ is good. The accuracy of our results is about
$20-30\%$, the main source of uncertainty is quark condensate
value and possible non-logarithmic perturbative corrections.  The
result of our calculation of the valence quark distribution
function (QDF) $x u_{\pi}(x,Q^2_0)$ in the pion is shown  in
Fig.1. Comparison with various experimental result lead us to
conclusion, that agreement with experiment is good. I want to
note, that this result of QDF is based only on the QCD sum rules.

If we use some natural additional assumptions, we also can
calculate the second moments of QDF.  Assume, that $u_{\pi}(x)
\sim 1/\sqrt{x}$ at small $x \la 0.15$ according to the Regge
behaviour and $u_{\pi}(x) \sim (1-x)^2$ at large $x \ga 0.7$
according to quark counting rules. Then, matching these functions
with uor result, one may find the numerical values of the first
and the second moments of the $u$-quark distribution: ${\cal{M}}_1
= \int \limits^1 _0 u_{\pi} (x) dx \approx 0.84$, ${\cal{M}}_2 =
\int \limits^1_0 xu_{\pi} (x) dx \approx 0.21$ where the values
various slightly at reasonable changes of QDF behaviour at large
or small $x$. The moment ${\cal{M}}_1$ has the meaning of the
number of $u$ quarks in $\pi^+$ and it should be ${\cal{M}}_1 =
1$. The deviation of $M_1$ from 1 characterizes the accuracy of
our calculation. The moment ${\cal{M}}_2$ has the meaning of the
pion momentum fraction carried by the valence $u$ quark.
Therefore, the valence $u$ and $\bar{d}$ quarks carry about 40\%
of the total momentum what is close to experimental results.

In the same way one can calculate valence $u$-quark distribution
in the $\rho^+$ meson. The choice of hadronic current is evident.
$j_{\mu}^{\rho} = \overline{u}\gamma_{\mu}d$ and the matrix
element is $\langle \rho^+\mid j^{\rho}_{\mu}\mid 0 \rangle =
\frac{m^2_{\rho}}{g_{\rho}}e_{\mu}$
 where $m_{\rho}$ is the $\rho$-meson mass, $g_{\rho}$ is the
$\rho-\gamma$ coupling constant, $g^2_{\rho}/4\pi=1.27$ and
$e_{\mu}$  is the $\rho$ meson polarization vector.

In the non-forward amplitude the tensor structure for
determination of $u$-quark distribution in the longitudinal $\rho$
meson is to $P_{\mu} P_{\nu} P_{\sigma}P_{\lambda}$, while for
transverse $\rho$ it is $-P_{\mu}P_{\nu}\delta_{\lambda\sigma}$)
(see \cite{3}).

In the case of longitudinal $\rho$ meson the tensor structure,
that is separated  is the same as in the case of the pion. It was
shown, that $u$-quark distribution in the longitudinal $\rho$
meson can be found from Eq.(1) by substituting $m_{\pi}\to
m_{\rho}$, $f_{\pi}\to m_{\rho}/g_{\rho}$ Sum rules for
$u^L_{\rho}(x)$  are satisfied in the wide $x$ region:  $0.1 < x <
0.85$ with high accuracy (about $10\%$). Figure. 1 (curve with
squares) presents $xu^L_{\rho}(x)$ as a function of $x$. The
values $M^2=1$ GeV$^2$ and $s_0=1.5$ GeV$^2$, $Q_0^2=4$ GeV$^2$
were chosen.

Let us now consider the case of transverse $\rho$-meson, i.e., the
term proportional to the structure
$P_{\mu}P_{\nu}\delta_{\lambda\sigma}$. The procedure of
calculations are the same except two points.

1. In contrast to the pion case, the $\langle
G^a_{\mu\nu}G^a_{\mu\nu}\rangle$ correction for transversally
polarized $\rho(\rho_T)$, does not vanish.

2. In contrast to $\pi$ and $\rho_L$-meson cases, the terms
proportional to\\ $\langle 0\mid  g^3 f^{abc}
G^a_{\mu\nu}G^b_{\nu\rho}$ $G^c_{\rho\mu} \mid 0 \rangle$  are not
cancelled for $\rho_T$ and one should estimate it. But $\langle 0
\mid g^3 G^a_{\mu\nu} G^b_{\nu\rho} G^c_{\rho\mu}f^{abc} \mid 0
\rangle$ is not well known; so we need to use estimations based
there are only some estimates based on the instanton model. Result
for QDF in transverasally polarized $\rho$ meson is shown on Fig.1
for the region $0.2<x<0.65$,(dash line). One can find complete
analytical form and it detail analysis in our paper \cite{3}, I do
not write it down since it is very large. The choose of parameters
are the same as in previous cases. The main sources of
uncertainities are $d=6$ gluon condensate (gives about $20\%$) and
$d=4$ gluon condensate (due the uncertainity of it own value about
factor 1.5, which lead to uncertainty in QDF about $20\%$). We
estimate accuracy about $30-50\%$ (better in the middle of region
and worser at the end). From fig.1 one can see, that difference
between QDF for longitudally $\rho_L$ and $\rho_T$ transversally
$\rho_T$ polarized $\rho$ meson is very large, many times larger
that uncertainties.

So we can conclude that QDF significantly depend of polarization.

Moreover, if we construct the moment of QDF for $\rho_L$ in the
same way as for pion then we found $M_1=1.06$ and $M_2=0.39$ (for
one valence quark). So we see that valence quark carry about
$80\%$ of the total momentum  of longitudinally polarized $\rho$
meson, so gluon sea there should be strobngly suppressed. One
should note, that accuracy of this prediction of moment of
$\rho_L$ is very high, because in this case sum rules for QDF are
covered almost all $x$ region, as I noted before, so extrapolation
procedure contribution numerically is negligible.
\section*{Acknowledgements} This study was supported in part
 by the Russian Foundation of Basic Research, project no.
RUP2 2621 MO-04 and by INTAS Call 2000, project 587.

\end{document}